\documentclass[superscriptaddress,preprintnumbers,amsmath,amssymb,twocolumn,floats,aps,prl]{revtex4-2}
\usepackage{txfonts}
\usepackage{amssymb}
\usepackage{graphicx}
\usepackage{epstopdf}
\usepackage{sidecap}
\usepackage{CJK}
\usepackage{txfonts}
\usepackage{url}
\usepackage{color}
\usepackage{indentfirst}
\usepackage{setspace}
\DeclareUnicodeCharacter{2061}{}
\begin{document}

\title{Kagome surface states and weak electronic correlation in vanadium-kagome metals}

\author{Jianyang Ding}
\thanks{Equal contributions}
%%\email{dingjy@mail.sim.ac.cn}
\affiliation{National Key Laboratory of Materials for Integrated Circuits, Shanghai Institute of Microsystem and Information Technology, Chinese Academy of Sciences, Shanghai 200050, China}
\affiliation{Center of Materials Science and Optoelectronics Engineering, University of Chinese Academy of Sciences, Beijing 100049, China}

\author{Ningning Zhao}
\thanks{Equal contributions}
%%\email{2015101973@ruc.edu.cn}
\affiliation{Department of Physics and Beijing Key Laboratory of Opto-Electronic Functional Materials $\&$ Micro-Nano Devices, Renmin University of China, Beijing 100872, China}

\author{Zicheng Tao}
\thanks{Equal contributions}
%%\email{taozch@shanghaitech.edu.cn}
\affiliation{School of Physical Science and Technology, ShanghaiTech University, Shanghai 201210, China}

\author{Zhe Huang}
%%\email{huangzhe@shanghaitech.edu.cn}
\affiliation{National Key Laboratory of Materials for Integrated Circuits, Shanghai Institute of Microsystem and Information Technology, Chinese Academy of Sciences, Shanghai 200050, China}
\affiliation{School of Physical Science and Technology, ShanghaiTech University, Shanghai 201210, China}

\author{Zhicheng Jiang}
%%\email{jiangzhch@mail.sim.ac.cn}
\affiliation{National Key Laboratory of Materials for Integrated Circuits, Shanghai Institute of Microsystem and Information Technology, Chinese Academy of Sciences, Shanghai 200050, China}
\affiliation{Center of Materials Science and Optoelectronics Engineering, University of Chinese Academy of Sciences, Beijing 100049, China}

\author{Yichen Yang}
%%\email{yangych@mail.sim.ac.cn}
\affiliation{National Key Laboratory of Materials for Integrated Circuits, Shanghai Institute of Microsystem and Information Technology, Chinese Academy of Sciences, Shanghai 200050, China}
\affiliation{Center of Materials Science and Optoelectronics Engineering, University of Chinese Academy of Sciences, Beijing 100049, China}

\author{Soohyun Cho}
%%\email{scho35@outlook.com}
\affiliation{National Key Laboratory of Materials for Integrated Circuits, Shanghai Institute of Microsystem and Information Technology, Chinese Academy of Sciences, Shanghai 200050, China}

\author{Zhengtai Liu}
%%\email{ztliu@mail.sim.ac.cn}
\affiliation{National Key Laboratory of Materials for Integrated Circuits, Shanghai Institute of Microsystem and Information Technology, Chinese Academy of Sciences, Shanghai 200050, China}
\affiliation{Center of Materials Science and Optoelectronics Engineering, University of Chinese Academy of Sciences, Beijing 100049, China}

\author{Jishan Liu}
%%\email{jishanliu@mail.sim.ac.cn}
\affiliation{National Key Laboratory of Materials for Integrated Circuits, Shanghai Institute of Microsystem and Information Technology, Chinese Academy of Sciences, Shanghai 200050, China}
\affiliation{Center of Materials Science and Optoelectronics Engineering, University of Chinese Academy of Sciences, Beijing 100049, China}

\author{Yanfeng Guo}
\email{guoyf@shanghaitech.edu.cn}
\affiliation{School of Physical Science and Technology, ShanghaiTech University, Shanghai 201210, China}

\author{Kai Liu}
\email{kliu@ruc.edu.cn}
\affiliation{Department of Physics and Beijing Key Laboratory of Opto-Electronic Functional Materials $\&$ Micro-Nano Devices, Renmin University of China, Beijing 100872, China}

\author{Zhonghao Liu}
\email{lzh17@mail.sim.ac.cn}
\affiliation{National Key Laboratory of Materials for Integrated Circuits, Shanghai Institute of Microsystem and Information Technology, Chinese Academy of Sciences, Shanghai 200050, China}
\affiliation{Center of Materials Science and Optoelectronics Engineering, University of Chinese Academy of Sciences, Beijing 100049, China}

\author{Dawei Shen}
\email{dwshen@mail.sim.ac.cn}
\affiliation{National Key Laboratory of Materials for Integrated Circuits, Shanghai Institute of Microsystem and Information Technology, Chinese Academy of Sciences, Shanghai 200050, China}
\affiliation{Center of Materials Science and Optoelectronics Engineering, University of Chinese Academy of Sciences, Beijing 100049, China}

\begin{abstract}
$R$V${_6}$Sn${_6}$ ($R$ = Y and lanthanides) with two-dimensional vanadium-kagome surface states is an ideal platform to investigate kagome physics and manipulate the kagome features to realize novel phenomena. Utilizing the micron-scale spatially resolved angle-resolved photoemission spectroscopy and first-principles calculations, we report a systematical study of the electronic structures of $R$V${_6}$Sn${_6}$ ($R$ = Gd, Tb, and Lu) on the two cleaved surfaces, i.e., the V- and $R$Sn$_1$-terminated (001) surfaces. The calculated bands without any renormalization match well with the main ARPES dispersive features, indicating the weak electronic correlation in this system. We observe '$W$'-like kagome surface states around the Brillouin zone corners showing $R$-element-dependent intensities, which is probably due to various coupling strengths between V and $R$Sn$_1$ layers. Our finding suggests an avenue for tuning electronic states by interlayer coupling based on two-dimensional kagome lattices.
\end{abstract}

\maketitle

\noindent\textbf{1. Introduction}\\

Kagome-lattice materials, owing to the geometric frustration, are ideal platforms to explore frustrating phenomena, electronic correlations, and quantum topology. Typical kagome-electronic bands feature Dirac-like dispersions at the Brillouin zone (BZ) corners, saddle points at the zone boundaries, and flat bands through the BZ~\cite{CoSn_Liu}. Recently, layered kagome-lattice 3$d$-transition-metal compounds have been discovered to host abundant quantum phenomena associating with the features near the Fermi energy ($E\rm_F$), such as Dirac and Weyl fermions ~\cite{Mn3Sn_Kuroda,Fe3Sn2_Linda,FeGeTe_Kim,CoSnS_LIU,CoSnS_Wang,CoMnGa_Hasan,CoSnS_YLChen,CoSnS_Beidenkopf,FeSn_Kang,FeSn_Lin,YMnSn_Li}, ferromagnetism ~\cite{malaman,Fe3Sn2_Lin,Fe3Sn2_Yin,FeGeTe_Zhang}, negative flat-band magnetism~\cite{CoSnS_Yin}, charge-density-wave states and possible unconventional superconductivity~\cite{Cs135_wilson,Rb135_Liu,Rb135_Shen}.

Among them, $R$V$_6$Sn$_6$ and $R$Mn$_6$Sn$_6$ ($R$ = Y and lanthanides) with ideal two-dimensional (2D) V and Mn kagome-lattice respectively have generated great interest. $R$Mn$_6$Sn$_6$ has been widely studied for its exhibiting a fascinating variety of magnetic ground states depending on different $R$ elements ~\cite{YMnSn_Li,GdMnSn_Liu,TbMnSn_Yin,R166_SJia}.
It has been reported that Chern-gapped topological fermions proposed in the spinless Haldane model could be realized in TbMn$_6$Sn$_6$, which hosts an out-of-plane ferromagnetic order~\cite{TbMnSn_Yin}. Typical kagome electronic bands have been observed in helical antiferromagnet YMn$_6$Sn$_6$~\cite{YMnSn_Li} and ferrimagnet GdMn$_6$Sn$_6$~\cite{GdMnSn_Liu}. The linear Dirac dispersion near $E\rm_F$ with intrinsic Berry curvature can induce the anomalous Hall effect and quantum oscillations ~\cite{YMnSn_Li,Gd166_Ronning,R166_SJia}. The calculated bands have to be renormalized by a factor of $\sim$ 2 to match the main observed dispersions, indicative of the moderate electron correlations in $R$Mn$_6$Sn$_6$~\cite{GdMnSn_Liu,R166_Gu}. On the other hand, although 4$f$ electrons of $R$ atoms could not affect bands near $E\rm_F$~\cite{GdMnSn_Liu}, the 3$d$ electron magnetism and correlations would be affected by 4$f$ electrons via the magnetic exchange coupling ~\cite{Tb166_connor,R166_SJia}. However, due to the electronic correlations and the magnetic domains/orders in $R$Mn$_6$Sn$_6$, some observations in this system are complicated and hard to analyze.

$R$V$_6$Sn$_6$ is isostructural to $R$Mn$_6$Sn$_6$, and nonmagnetic V atoms form ideal 2D kagome-lattice in $R$V$_6$Sn$_6$. Without magnetic orders and magnetic domains in V-kagome layers, $R$V$_6$Sn$_6$ would be advantageous for clearly addressing fundamental issues of kagome physics both theoretically and experimentally. Moreover, the $R$ elements in the adjacent layers cannot change the main electronic structure near $E\rm_F$ defined by V 3$d$ orbitals but can supply various magnetic moments which could modify kagome physics in the V layers. Recent studies of $R$V$_6$Sn$_6$ have revealed kagome surface states and topological surface states~\cite{V166_JFHe,V166_MShi,TbVS,V166_NP}. ScV$_6$Sn$_6$ has been extensively studied as the only compound among the series of $R$V$_6$Sn$_6$ that displays a charge density wave (CDW) order~\cite{ScVS_1,ScVS_2,ScVS_3,ScVS_4,ScVS_5,ScVS_6,ScVS_7,ScVS_8}. However, whether kagome surface states can be affected by different $R$ elements has not been systematically studied.

In this work, we systematically study the electronic structures of $R$V${_6}$Sn${_6}$ ($R$ = Gd, Tb, and Lu) with two typical cleavage surfaces, i.e., the V- and $R$Sn$_1$-terminated (001) surfaces, utilizing the micron-scale spatially resolved angle-resolved photoemission spectroscopy ($\mu$ARPES) and the first-principles calculations. The observed band structure can be well captured by the calculations without any renormalization, indicating a weak electron correlation. We have unveiled typical bulk states and surface states of V-kagome lattices in all three compounds.  Especially, the intensity of surface states at $E\rm_F$ by normalized photocurrents in LuV${_6}$Sn${_6}$ is much stronger than the others, possibly because Lu (4$f^{14}$5$d^1$6$s^2$) with fulfilled 4$f$-shell has a weak interlayer coupling and less interference with surface states in the adjacent V-layer. Our findings suggest that kagome surface states at $E\rm_F$ in $R$V$_6$Sn$_6$ could be tuned by the interlayer coupling via various $R$ elements.
\\

\begin{figure}[!tbp]
%\centering
\includegraphics[width=9cm,scale=2.00]{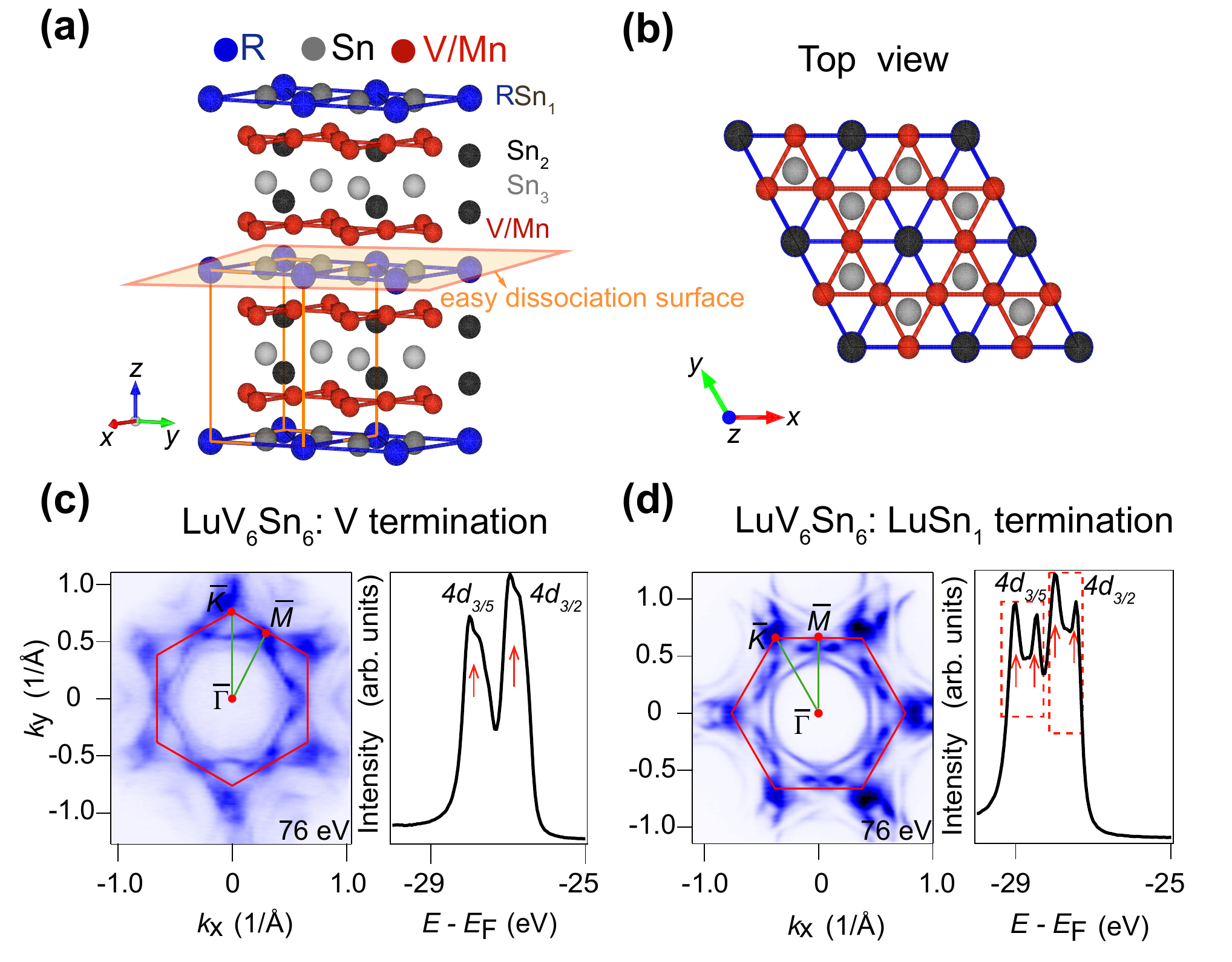}
\caption
{
  (a) Crystal structure of $R$V$_6$Sn$_6$. The unit cell and easily cleaved surface are marked with orange lines and plane, respectively.
  (b) Top view of the crystal structure along the $c$ axis shows the V kagome lattice and the projected Sn$_2$ and Sn$_3$ sites. 
  (c), (d) The Fermi surfaces and corresponding core-level spectra on the V and LuSn$_1$ terminations of LuV$_6$Sn$_6$, respectively.
}
\label{1}
\end{figure}

\noindent\textbf{2. Methods}\\

The $\mu$ARPES measurements were performed at the BL-03U beamline of the Shanghai Synchrotron Radiation Facility (SSRF)~\cite{sun2020_03U,yang2021_03U}. The energy and angular resolutions were set to better than 20 meV and 0.02 {\AA}$^{-1}$, respectively. The light spot size is smaller than 20 $\mu$m. Samples were cleaved $in$ $situ$, exposing flat mirrorlike (001) surfaces. During measurements, the temperature was kept at T $\sim$ 15 K, and the pressure was maintained at less than $8\times10^{-11}$ Torr. The used photon energies are from 50 to 100 eV.

The first-principles calculations were performed by using the projector augmented wave (PAW) method~\cite{DFT_1,DFT_2} as implemented in the Vienna $ab$ $initio$ simulation package (VASP)~\cite{DFT_3,DFT_4,DFT_5}. The generalized gradient approximation (GGA) of Perdew-Burke-Ernzerhof (PBE) type~\cite{DFT_6} was used for the exchange-correlation functional. The valence configurations for Lu, V, and Sn atoms are $5p^{6}5d^{1}6s^{2}$, $3s^{2}3p^{6}3d^{4}4s^{1}$, and $4d^{10}5s^{2}5p^{2}$, respectively. The lattice constants were fixed to the experimental values of a = b = 5.5348 $\AA$, c = 9.1797$\AA$ for GdV$_6$Sn$_6$ and a = b= 5.5039 $\AA$, c = 9.1764 $\AA$ for LuV$_6$Sn$_6$ [4]. The spin-orbit coupling (SOC) effect was included in the band structure calculations. The kinetic energy cutoff of the plane-wave basis was set to 350 eV. The BZ was sampled with an 11 × 11 × 7 $k$-point mesh. For the Fermi surface broadening, the Gaussian smearing method with a width of 0.05 eV was adopted.  The surface states for V kagome and $R$Sn$_1$ terminations in the projected 2D BZ were calculated with the surface Green’s function method by using the WannierTools package~\cite{DFT_7}. The tight-binding Hamiltonian of the semi-infinite system was constructed by the maximally localized Wannier functions for the outmost $s$, $p$, $d$, and $f$ orbitals of Lu atoms, $s$, $p$, and $d$ orbitals of V atoms and $s$, $p$ and $d$ orbitals of Sn atoms generated by the first-principles calculations.~\cite{DFT_8,DFT_9}.
\\

\noindent\textbf{3. Results and analyses}\\

Based on the $\mu$ARPES observations and the first-principles calculations, the surface states have been distinguished from the bulk states on the two different cleaved surfaces. The '$W$'-like kagome surface states around the BZ corners show $R$-element-dependent intensities, and the calculated bulk bands without any renormalization match well with the main ARPES dispersive features. We give in detail the results and analyses below.

Single crystals of $R$V$_6$Sn$_6$ were synthesized by the self-flux method~\cite{R166_romaka,R166_lee}. $R$V$_6$Sn$_6$ crystallizes in the hexagonal HfFe$_6$Ge$_6$-type structure with the space group $\emph{P6/mmm}$ (No. 191). The three-dimensional (3D) crystal structure of $R$V$_6$Sn$_6$ are built by stacking of $R$Sn$_1$-VSn$_2$-Sn$_3$-VSn$_2$-$R$Sn$_1$ along the $c$ axis, as shown in Fig.~\ref{1}(a). V atoms without magnetic moment constitute kagome lattices and $R$ atoms with a magnetic moment in the adjacent layer are arranged in triangular lattices centered by Sn$_1$ atoms [Fig.~\ref{1}(b)]. Considering the chemical bonding energy and bond length~\cite{T166_pokharel,V166_MShi}, the sample is expected to be cleaved between neighboring V and $R$Sn$_1$ planes [Fig.~\ref{1}(a)]. In the practical case in experimental measurements, the cleavage surface of $R$V$_6$Sn$_6$ should be composed of multiple V and $R$Sn$_1$ domains, which have been reported to have distinct electronic structures ~\cite{V166_JFHe,V166_MShi}. 

It is crucial to pinpoint the single domain in the photoemission measurements. Utilizing the micron-scale spatially resolved ARPES, we performed a real-space map of photoelectrons on the cleaved surface. Figures~\ref{1}(c) and ~\ref{1}(d) present the Fermi surfaces and the corresponding core-level spectra taken on the V and LuSn$_1$ terminations of LuV$_6$Sn$_6$, respectively. The former presents relatively simple Fermi surfaces and Sn-4$d$ peaks without distinct splitting, while the latter shows a splitting Fermi surface and the two Sn-4$d$ side peaks. Detailed core-level spectra from different domains of LuV${_6}$Sn${_6}$ are compared in Figs.~S1(a) and S1(b)~\cite{SI_Here}. The spectra taken on the V termination illustrate the weaker intensity of Lu-4$f$ peaks and less splitting of Sn-4$d$ peaks, while those from LuSn$_1$ domains are just the opposite, consistent with the previous report~\cite{V166_JFHe}.

\begin{figure}[!htbp]
%\centering
\includegraphics[width=9cm,scale=2.00]{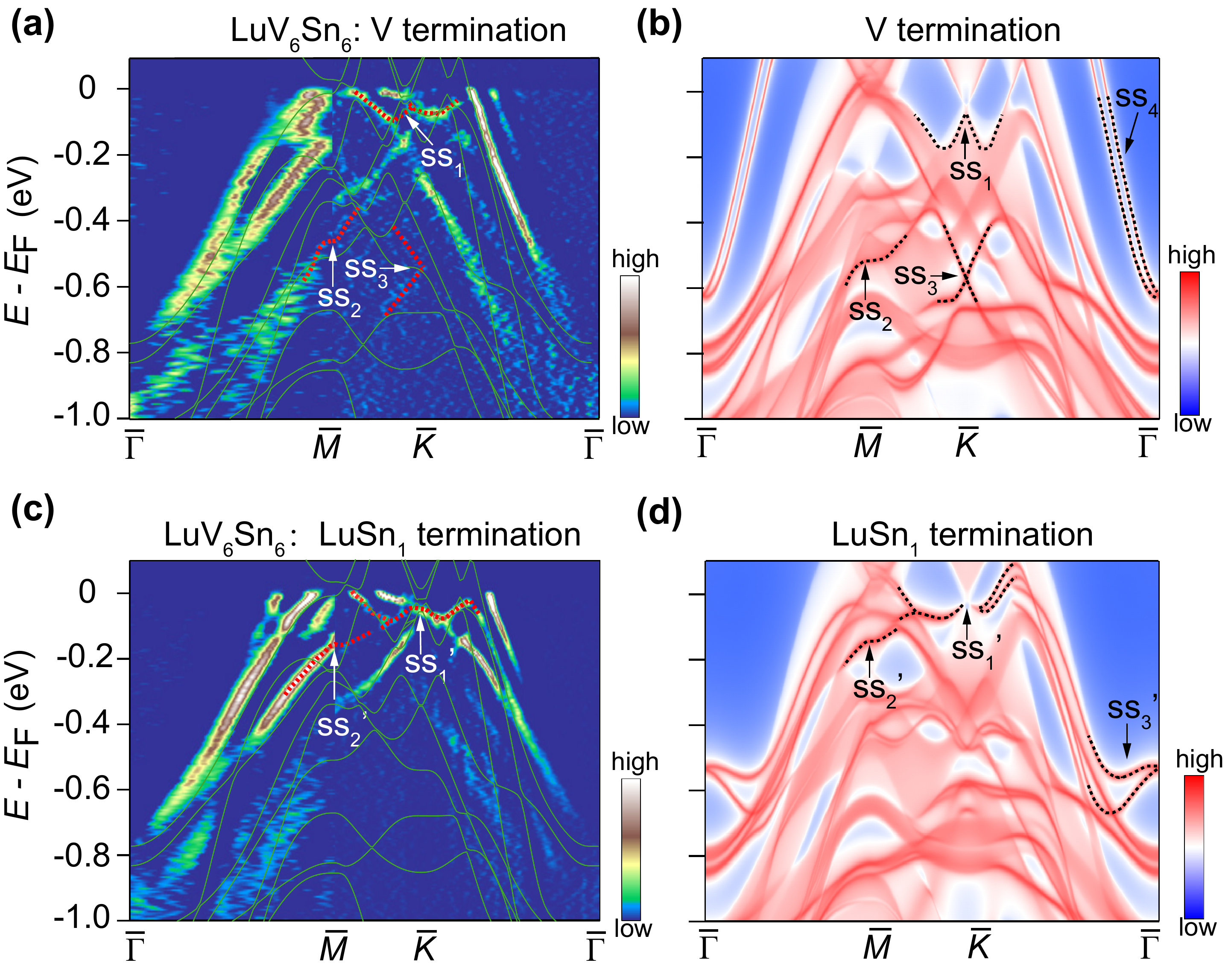}
\caption
{
 (a), (b) The second-derivative plots of the measured bands and the projected surface states along the high symmetry directions ($\overline{\Gamma}$-$\overline{K}$-$\overline{M}$-$\overline{\Gamma}$) for $V$-terminated (001) surfaces of LuV$_6$Sn$_6$. 
 (c), (d) The same as in (a), (b), but for $R$Sn$_1$-terminated (001) surfaces. The green solid lines in (a) and (c) are the calculated bulk bands, and the dotted lines are indicated the surface states.
}
\label{2}
\end{figure}

\begin{figure*}[!htbp]
\centering
\includegraphics[width=16cm]{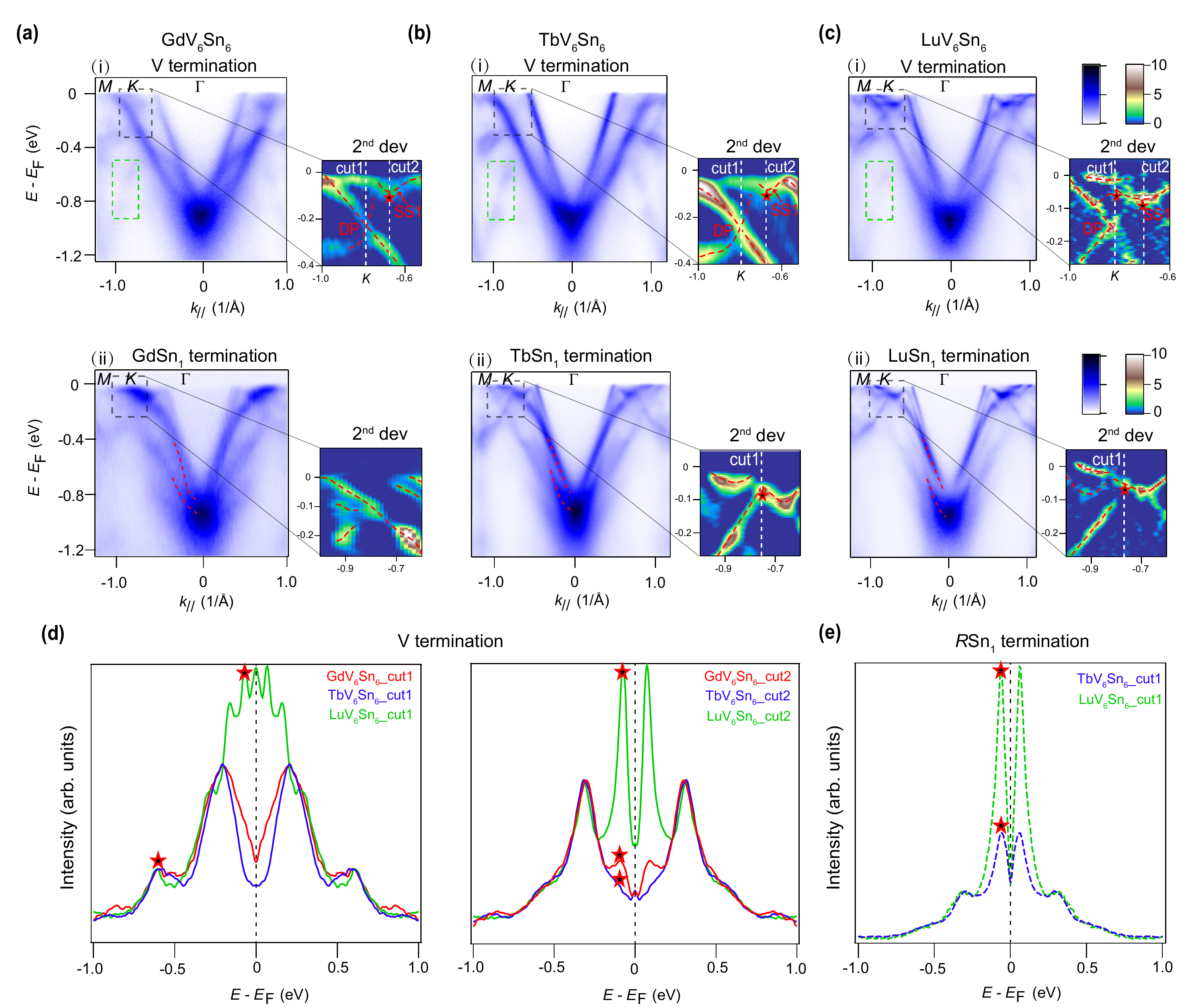}
\caption
{ 
 (a)-(c) The first row (i) is the band dispersions along the $\overline{\Gamma}$-$\overline{\emph{K}}$-$\overline{\emph{M}}$ directions on the V terminations of $R$V$_6$Sn$_6$, taken with 76-eV photons. The second-derivative plots correspond to enlarged parts as indicated by the black boxes. The red dashed lines indicate dispersions around $K$ point near $E\rm_F$. The green boxes indicate the Dirac surface state dispersions at about -0.6 eV below $E\rm_F$. The second row (ii) is the same as in (i), but taken on the $R$Sn$_1$ terminations. All the ARPES data are normalized in the same condition to rule out the change of photocurrent intensities induced by uncertainty factors in the measurements. 
 (d), (e) The symmetrized EDCs at cut1 and cut2 in the second-derivative plots remove the effect of the Fermi-Dirac function showing the band tops and bottoms, respectively. The asterisks mark EDCs peaks of surface state (SS1) at cut1 and cut2, and the squares mark the Dirac cone positions. 
}
\label{3}   
\end{figure*}

After identifying spectra taken from different domains, we explored the features of the kagome surface states on different terminations. Figure~\ref{2} demonstrates comparisons between the measured and calculated surface states along the $\overline{\Gamma}$-$\overline{M}$-$\overline{K}$-$\overline{\Gamma}$ direction for V- and LuSn$_1$-terminated (001) surfaces of LuV$_6$Sn$_6$. The calculated bulk bands have been appended on the corresponding photoemission intensity plots in Figs.~\ref{2}(a) and ~\ref{2}(c). There exist several bands escaping from the calculated bulk band structure, as indicated by the red dotted lines and the white arrows. These bands do not disperse along the $k_z$ direction, as shown in Figs.~S4 and S5~\cite{SI_Here}. By further comparing with calculated surface bands in Figs.~\ref{2}(b) and ~\ref{2}(d), we can determine that they should originate from the surface states. At the $\overline{K}$ point, a '$W$'-like surface state (SS1) is located around $E\rm_F$, and a Dirac surface state (SS3) on the V termination is located at about -0.6 eV below $E\rm_F$. At the $\overline{M}$ point, the saddlelike surface states (SS2) are located at about -0.5 eV below $E\rm_F$ on the V termination and at about -0.2 eV below $E\rm_F$ on the LuSn$_1$ termination (SS2$^\prime$), respectively. The observed kagome surface states are in line with the previous report~\cite{V166_JFHe}. At the $\overline{\Gamma}$ point, although the calculation suggests the existence of some surface states (SS4 on the V termination and SS3$^\prime$ on the LuSn$_1$ termination), these states are invisible in the photoemission results, which might be due to the mixture of the bulk states and the matrix element effects [Figs.~S4(a) and ~S4(b)]~\cite{SI_Here}.

\begin{figure*}[!htbp]
\centering
\includegraphics[width=16.5cm]{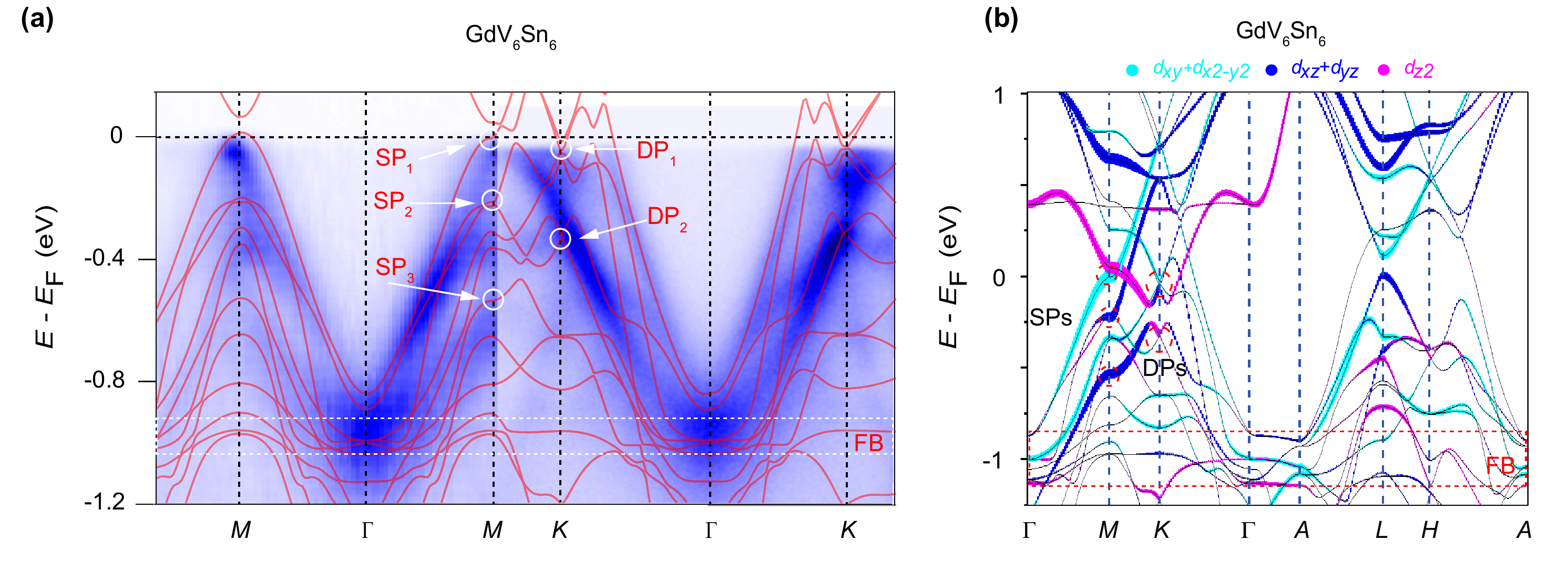}
\caption
{
  (a) The ARPES intensity plot of GdV$_6$Sn$_6$ along the high symmetry directions, taken with 82-eV photons. The calculated bands with the spin-orbit coupling (SOC) for the paramagnetic state are appended, as shown in the red solid lines. 
  (b) Orbital-projection band-structure calculation of GdV$_6$Sn$_6$ with SOC for the paramagnetic state. The orbital weights are represented by both the colors and the size of the bands. The saddle points (SPs), Dirac points (DPs), and Flat band (FB) are indicated.
}
\label{4}   
\end{figure*}

To investigate the effect of different lanthanides elements on the electronic structure of $R$V$_6$Sn$_6$, we systematically performed the high-resolution ARPES measurements on both the V- and $R$Sn$_1$-terminated (001) surfaces of $R$V$_6$Sn$_6$ family. Figures~\ref{3}(a)-~\ref{3}(c) side-by-side compare experimental bands of $R$V$_6$Sn$_6$ along the $\overline{\Gamma}$-$\overline{K}$-$\overline{M}$ directions on the two different terminations, i.e., (i) the V terminations in the first row, and (ii) the $R$Sn$_1$ terminations in the second row. 
Here, the same experimental setup has been used in all ARPES measurements, namely the samples measured under the same photon energy with the same polarization and flux, the same geometry and temperature, etc. All collected data have been normalized by the background above $E\rm_F$ to get rid of the extrinsic interference as shown in Fig. S4~\cite{SI_Here}.
In this way, the bulk band structures of different $R$-element compounds do not exhibit remarkable changes globally for the same termination. However, we noticed that the dispersions obtained from different terminations show a pronounced difference, especially for electronlike bands around the $\Gamma$ point, which indicates that the bulk states strongly mix with the surface states at $\Gamma$ on the $R$Sn$_1$ terminations as shown in Fig.~\ref{2}(d). 

In all three compounds, the Dirac surface states located at about -0.6 eV below $E\rm_F$ [indicated by the green boxes in Figs.~\ref{3}(a)-~\ref{3}(c) and Fig.~S4(e)] can only be observed on the V terminations ~\cite{SI_Here}, but not on the $R$Sn$_1$ terminations, which is also in line with our calculation shown in Figs.~\ref{2}(b) and \ref{2}(d). Here only one branch of the Dirac surface state can be clearly observed due to the matrix element effects associated with the chirality of the Dirac fermion~\cite{CoSn_kang}. According to the calculation in Fig.~\ref{2} and the previous study of HoV$_6$Sn$_6$~\cite{V166_JFHe}, we could identify a '$W$'-like band at the $K$ point as the 2D-kagome surface state, as shown in the second-derivative plot in Fig.~\ref{3}.

To analyze in detail the effect of the $R$ elements on the surface states, we extracted the energy distribution curves (EDCs) along cut1 (at the $K$ point) and cut2 ($k\rm_{//}$ = 0.7 $\AA^{-1}$) as indicated in the corresponding second-derivative plots, and then symmetrized them with respect to $E\rm_F$ which is expected to approximately remove the effect of the Fermi-Dirac function cutoff~\cite{NFCA_liu,liu_AlFeB}. As shown in Figs.~\ref{3}(d) and \ref{3}(e), these symmetrized EDCs look rather similar at higher binding energy, because different intensities in bulk states are approximately eliminated after the normalization. The Dirac surface states at -0.6 eV below $E\rm_F$ with almost the same intensity, marked by the asterisks, are observed only on the V termination in all three materials [Fig.~\ref{3}(d) and Fig. S6(e)]. In sharp contrast, in the vicinity of $E\rm_F$, the spectral intensity is seriously modified by the '$W$'-like kagome surface state on both terminations [the asterisks in Figs.~\ref{3}(d) and ~\ref{3}(e)], and the intensity of the '$W$'-like surface state is the most enhanced in LuV$_6$Sn$_6$.  

To study electronic correlations of $R$V$_6$Sn$_6$, we compared the calculation with ARPES data in the paramagnetic states [Fig.~\ref{4}(a)], which illustrates good overall agreement between them without any renormalization. The typical kagome bulk bands are found here: the saddle points at $M$, the Dirac cones at $K$, and the flat band running through the BZ. 
The Dirac cone near $E\rm_F$ (DP1) forming a two-gap feature by SOC and the observed flat band are in agreement with the recent studies in $R$V$_6$Sn$_6$ compounds~\cite{V166_NP}.
Compared with GdMn$_6$Sn$_6$~\cite{GdMnSn_Liu}, which is a correlated Hund's metal reflected by the Hubbard $U$ and the renormalization factor, GdV$_6$Sn$_6$ with nonmagnetic V atoms and weak electronic correlation is easy to be studied by the calculations and experiments, making it a simple platform to study fundamental kagome physics, such as the 2D surface states mentioned above. 

Figure~\ref{4}(b) shows the orbital projected band structure of GdV$_6$Sn$_6$. The bands near $E\rm_F$ are mainly contributed by the V 3$d$ orbitals, and they are strongly hybridized. With the help of the calculation, one can find that the saddle points at $M$ are mainly derived from the $d_{xy}$+$d{_{x^{2}-y^{2}}}$ and $d_{xz}$+$d_{yz}$ orbitals, and the Dirac dispersions at $K$ are mainly originated from $d_{z^{2}}$, $d_{xz}$ and $d_{yz}$ orbitals. 
For the flat band, the interlayer hybridization associated with the multi $d$ orbitals is likely to disturb the  destructive interference of Bloch electrons, leading to fragmented flatness observed at around -1 eV below $E\rm_F$, as in Fe$_3$Sn$_2$~\cite{Fe3Sn2_Lin} and $R$Mn$_6$Sn$_6$~\cite{YMnSn_Li,GdMnSn_Liu} etc. Additionally, the hybridized flat bands with complicated matrix element effects could also bring the difficulty of the ARPES measurements~\cite{PPS_park,Y166_yang}.
\\

\noindent\textbf{4. Discussions}\\

Some possible factors should be considered in the ARPES spectral intensity, such as the experimental setup, the sample surfaces, element-dependent cross-sections, and matrix element effects. Thus, the same experimental setup has been used in the ARPES measurements, and the collected data have been normalized in the same way as mentioned above. In this way, the intensities of the corresponding bulk bands are almost the same in the three compounds [Figs.~\ref{3}(d) and ~\ref{3}(e)]. While the intensities of '$W$'-like surface states show noticeable changes. 

On measuring the various samples (Fig. S8)~\cite{SI_Here}, we find that the surface state in LuV${_6}$Sn${_6}$ is much stronger than that in the other compounds. Since the surface state is more sensitive to the cleaved surfaces than the bulk state, we cannot rule out the factor of the cleaved surfaces entirely. Based on the consistency of our plenteous data showing $R$-dependent intensities of surface states, we try to propose a reasonable explanation in physics as follows. 

A previous report reveals that the lattice constants (both $a$ and $c$) and magnetic ordering temperature (T$_N$) of $R$V$_6$Sn$_6$ would be decreased with the increasing of $R$ atomic number, and the valence states of $R$ ions are trivalent~\cite{R166_lee}. The de Gennes factor of $R^{3+}$, ${dG=(g{_J}-1)^{2}J(J+1)}$, where $g{_J}$ is the Land\'{e} factor and $J$ is the total angular momentum of the Hund's rule ground state, and  which is related to the different exchange interaction caused by the 4$f$ electronic local moments of the $R^{3+}$ ions and interlayer electron hopping, becomes weaker from Gd$^{3+}$ to Lu$^{3+}$ (See the supplementary material TABLE I)~\cite{SI_Here}. Referring to the study of $R$Mn$_6$Sn$_6$ ~\cite{R166_SJia}, the electron hopping between interlayers would affect the band dispersions and topological properties. We consider the first-order perturbation, $H_{1}$=${-J_{H}m{\sum}{_{i}}c{_{i}}^{+}{c_{i}}}{\sim}-J{_{H}}{\sqrt{dG}}$, introduced by the interlayer electron hopping in the mean-field approximation, where $J\rm_H$ is the Hund's coupling. Along with the increasing $R$ atomic number, $R$V$_6$Sn$_6$ has a smaller interlayer electron hopping and $\left|H_{1}\right|$ to make the V-kagome layer less perturbed by the $R$Sn$_1$ layer. The interlayer electron hopping along the $z$ direction described by $H_1$ perturbation would mainly affect the stability of the electron occupation states on the 2D V-kagome layer, which may be quantitatively reflected in $A(k.w)$ (single particle excitation spectral function), but it is very difficult to accurately represent it quantitatively in terms of spectral intensity. Moreover, the band structure around the $K$ point is mainly of the $d_{z^{2}}$ orbital [Fig.~\ref{4}(b)], which may also be the reason for inducing electron hopping along $z$.

LuV$_6$Sn$_6$ with fulfilled 4$f$-shell (Lu 4$f^{14}$5$d^1$6$s^2$) shows the much strongest intensity of the '$W$'-like surface states at $E\rm_F$. The effect of the $R$ elements on the surface states could be limited at $E\rm_F$. At higher binding energy, as for the Dirac surface states at -0.6 eV below $E\rm_F$ with almost the same intensity, this effect could be neglected. The mechanism of the $R$ element tuned kagome surface states in the $R$V$_6$Sn$_6$ system deserves to be further studied theoretically and experimentally.
\\

\noindent\textbf{5. Conclusions}\\

In summary, we systematically report the electronic structures of $R$V${_6}$Sn${_6}$ on the two typical cleaved (001) surfaces, finding the 2D kagome surface states and weak electronic correlation effect.
We give a possible reason that LuV$_6$Sn$_6$ with fulfilled 4$f$-shell showing the clearest surface states near $E\rm_F$ in the system. Our studies could stimulate further studies of tunable kagome physics of $R$V${_6}$Sn${_6}$ by the interlayer coupling.
\\

\noindent\textbf{Data availability statement}\\

The data that support the findings of this study are available upon request from the authors.
\\

\noindent\textbf{Acknowledgments}\\

This work was supported by the National Key R$\&$D Program of China (Grants No. 2022YFB3608000 and 2017YFA0302903), the National Natural Science Foundation of China (NSFC, Grants No. U2032208, 12222413, and 12174443), the Shanghai Science and Technology Innovation Action Plan (Grant No. 21JC1402000), the Natural Science Foundation of Shanghai (Grants No. 23ZR1482200 and 22ZR1473300), the Beijing Natural Science Foundation (Grant No. Z200005), and the CAS Interdisciplinary Innovation Team. N.Z. was supported by the Outstanding Innovative Talents Cultivation Funded Programs 2020 of Renmin University of China (RUC). Computational resources were provided by the Physical Laboratory of High-Performance Computing at RUC. J. S. L. thanks the fund of Science and Technology on Surface Physics and Chemistry Laboratory (Grant No. 6142A02200102). Part of this research used  the 03U Beamline of the Shanghai Synchrotron Radiation Facility, which is supported by the ME$^2$ project under Contract No. 11227902 from NSFC.

\bibliographystyle{apsrev4-2}
\bibliography{RV6Sn6}

\begin{thebibliography}{10}
\providecommand{\url}[1]{#1}
\csname url@samestyle\endcsname
\providecommand{\newblock}{\relax}
\providecommand{\bibinfo}[2]{#2}
\providecommand{\BIBentrySTDinterwordspacing}{\spaceskip=0pt\relax}
\providecommand{\BIBentryALTinterwordstretchfactor}{4}
\providecommand{\BIBentryALTinterwordspacing}{\spaceskip=\fontdimen2\font plus
\BIBentryALTinterwordstretchfactor\fontdimen3\font minus
  \fontdimen4\font\relax}
\providecommand{\BIBforeignlanguage}[2]{{%
\expandafter\ifx\csname l@#1\endcsname\relax
\typeout{** WARNING: IEEEtran.bst: No hyphenation pattern has been}%
\typeout{** loaded for the language `#1'. Using the pattern for}%
\typeout{** the default language instead.}%
\else
\language=\csname l@#1\endcsname
\fi
#2}}
\providecommand{\BIBdecl}{\relax}
\BIBdecl

\bibitem{SI_Here}
\emph{See {S}upplemental {M}aterials for additional data of the {ARPES}
  experiments and the first-principles calculations, which include {R}efs.
  \cite{sun2020_03U,yang2021_03U,DFT_1,DFT_2,DFT_3,DFT_4,DFT_5,DFT_6,DFT_7,DFT_8,DFT_9}}.

\bibitem{CoSn_Liu}
Z.~Liu, M.~Li, Q.~Wang, G.~Wang, C.~Wen, K.~Jiang, X.~Lu, S.~Yan, Y.~Huang,
  D.~Shen \emph{et~al.}, ``Orbital-selective {D}irac fermions and extremely
  flat bands in frustrated kagome-lattice metal {CoSn},'' \emph{Nature
  communications}, vol.~11, no.~1, p. 4002, 2020.

\bibitem{Mn3Sn_Kuroda}
K.~Kuroda, T.~Tomita, M.-T. Suzuki, C.~Bareille, A.~Nugroho, P.~Goswami,
  M.~Ochi, M.~Ikhlas, M.~Nakayama, S.~Akebi \emph{et~al.}, ``Evidence for
  magnetic {W}eyl fermions in a correlated metal,'' \emph{Nature Materials},
  vol.~16, no.~11, pp. 1090--1095, 2017.

\bibitem{Fe3Sn2_Linda}
L.~Ye, M.~Kang, J.~Liu, F.~Von~Cube, C.~R. Wicker, T.~Suzuki, C.~Jozwiak,
  A.~Bostwick, E.~Rotenberg, D.~C. Bell \emph{et~al.}, ``Massive {D}irac
  fermions in a ferromagnetic kagome metal,'' \emph{Nature}, vol. 555, no.
  7698, pp. 638--642, 2018.

\bibitem{FeGeTe_Kim}
D.~Kim, C.~Lee, B.~G. Jang, K.~Kim, and J.~H. Shim, ``Drastic change of
  magnetic anisotropy in {Fe$_3$GeTe$_2$} and {Fe$_4$GeTe$_2$} monolayers under
  electric field studied by density functional theory,'' \emph{Scientific
  Reports}, vol.~11, no.~1, p. 17567, 2021.

\bibitem{CoSnS_LIU}
E.~Liu, Y.~Sun, N.~Kumar, L.~Muechler, A.~Sun, L.~Jiao, S.-Y. Yang, D.~Liu,
  A.~Liang, Q.~Xu \emph{et~al.}, ``Giant anomalous {H}all effect in a
  ferromagnetic kagome-lattice semimetal,'' \emph{Nature Physics}, vol.~14,
  no.~11, pp. 1125--1131, 2018.

\bibitem{CoSnS_Wang}
Q.~Wang, Y.~Xu, R.~Lou, Z.~Liu, M.~Li, Y.~Huang, D.~Shen, H.~Weng, S.~Wang, and
  H.~Lei, ``Large intrinsic anomalous {H}all effect in half-metallic
  ferromagnet co$_3$sn$_2$s$_2$ with magnetic {W}eyl fermions,'' \emph{Nature
  communications}, vol.~9, no.~1, p. 3681, 2018.

\bibitem{CoMnGa_Hasan}
I.~Belopolski, K.~Manna, D.~Sanchez, G.~Chang, B.~Ernst, J.~Yin, S.~Zhang,
  T.~Cochran, N.~Shumiya, H.~Zheng, B.~Singh, G.~Bian, D.~Multer,
  M.~Litskevich, Z.~Xiaoting, S.-M. Huang, B.~Wang, T.-R. Chang, S.-Y. Xu,
  A.~Bansil, C.~Felser, H.~Lin, and Z.~Hasan, ``Discovery of topological {W}eyl
  fermion lines and drumhead surface states in a room temperature magnet,''
  \emph{Science}, vol. 365, no. 6459, pp. 1278--1281, 2019.

\bibitem{CoSnS_YLChen}
D.~Liu, A.~Liang, E.~Liu, Q.~Xu, Y.~Li, C.~Chen, D.~Pei, W.~Shi, S.~Mo,
  P.~Dudin \emph{et~al.}, ``Magnetic {W}eyl semimetal phase in a kagome
  crystal,'' \emph{Science}, vol. 365, no. 6459, pp. 1282--1285, 2019.

\bibitem{CoSnS_Beidenkopf}
N.~Morali, R.~Batabyal, P.~K. Nag, E.~Liu, Q.~Xu, Y.~Sun, B.~Yan, C.~Felser,
  N.~Avraham, and H.~Beidenkopf, ``Fermi-arc diversity on surface terminations
  of the magnetic {W}eyl semimetal {Co$_3$Sn$_2$S$_2$},'' \emph{Science}, vol.
  365, no. 6459, pp. 1286--1291, 2019.

\bibitem{FeSn_Kang}
M.~Kang, L.~Ye, S.~Fang, J.-S. You, A.~Levitan, M.~Han, J.~I. Facio,
  C.~Jozwiak, A.~Bostwick, E.~Rotenberg \emph{et~al.}, ``Dirac fermions and
  flat bands in the ideal kagome metal {FeSn},'' \emph{Nature Materials},
  vol.~19, no.~2, pp. 163--169, 2020.

\bibitem{FeSn_Lin}
Z.~Lin, C.~Wang, P.~Wang, S.~Yi, L.~Li, Q.~Zhang, Y.~Wang, Z.~Wang, H.~Huang,
  Y.~Sun \emph{et~al.}, ``Dirac fermions in antiferromagnetic {FeSn} kagome
  lattices with combined space inversion and time-reversal symmetry,''
  \emph{Physical Review B}, vol. 102, no.~15, p. 155103, 2020.

\bibitem{YMnSn_Li}
M.~Li, Q.~Wang, G.~Wang, Z.~Yuan, W.~Song, R.~Lou, Z.~Liu, Y.~Huang, Z.~Liu,
  H.~Lei, Z.~Yin, and S.~Wang, ``Dirac cone, flat band and saddle point in
  kagome magnet {YMn$_6$Sn$_6$},'' \emph{Nature Communications}, vol.~12, p.
  3129, 2021.

\bibitem{malaman}
B.~Malaman, G.~Venturini, R.~Welter, J.~Sanchez, P.~Vulliet, and E.~Ressouche,
  ``Magnetic properties of {$R$Mn$_6$Sn$_6$} ({R= Gd-Er}) compounds from
  neutron diffraction and m{\"o}ssbauer measurements,'' \emph{Journal of
  Magnetism and Magnetic Materials}, vol. 202, no. 2-3, pp. 519--534, 1999.

\bibitem{Fe3Sn2_Lin}
Z.~Lin, J.-H. Choi, Q.~Zhang, W.~Qin, S.~Yi, P.~Wang, L.~Li, Y.~Wang, H.~Zhang,
  Z.~Sun \emph{et~al.}, ``Flatbands and emergent ferromagnetic ordering in
  {Fe$_3$Sn$_2$} kagome lattices,'' \emph{Physical Review Letters}, vol. 121,
  no.~9, p. 096401, 2018.

\bibitem{Fe3Sn2_Yin}
J.-X. Yin, S.~S. Zhang, H.~Li, K.~Jiang, G.~Chang, B.~Zhang, B.~Lian, C.~Xiang,
  I.~Belopolski, H.~Zheng \emph{et~al.}, ``Giant and anisotropic many-body
  spin-orbit tunability in a strongly correlated kagome magnet,''
  \emph{Nature}, vol. 562, no. 7725, pp. 91--95, 2018.

\bibitem{FeGeTe_Zhang}
Y.~Deng, Y.~Yu, Y.~Song, J.~Zhang, N.~Z. Wang, Z.~Sun, Y.~Yi, Y.~Z. Wu, S.~Wu,
  J.~Zhu \emph{et~al.}, ``Gate-tunable room-temperature ferromagnetism in
  two-dimensional {Fe$_3$GeTe$_2$},'' \emph{Nature}, vol. 563, no. 7729, pp.
  94--99, 2018.

\bibitem{CoSnS_Yin}
J.-X. Yin, S.~S. Zhang, G.~Chang, Q.~Wang, S.~S. Tsirkin, Z.~Guguchia, B.~Lian,
  H.~Zhou, K.~Jiang, I.~Belopolski \emph{et~al.}, ``Negative flat band
  magnetism in a spin-orbit-coupled correlated kagome magnet,'' \emph{Nature
  Physics}, vol.~15, no.~5, pp. 443--448, 2019.

\bibitem{Cs135_wilson}
B.~R. Ortiz, S.~M. Teicher, Y.~Hu, J.~L. Zuo, P.~M. Sarte, E.~C. Schueller,
  A.~M. Abeykoon, M.~J. Krogstad, S.~Rosenkranz, R.~Osborn \emph{et~al.},
  ``{CsV$_3$Sb$_5$}: {A} {Z$_2$} topological kagome metal with a
  superconducting ground state,'' \emph{Physical Review Letters}, vol. 125,
  no.~24, p. 247002, 2020.

\bibitem{Rb135_Liu}
Z.~Liu, N.~Zhao, Q.~Yin, C.~Gong, Z.~Tu, M.~Li, W.~Song, Z.~Liu, D.~Shen,
  Y.~Huang \emph{et~al.}, ``Charge-density-wave-induced bands renormalization
  and energy gaps in a kagome superconductor {RbV$_3$Sb$_5$},'' \emph{Physical
  Review X}, vol.~11, no.~4, p. 041010, 2021.

\bibitem{Rb135_Shen}
S.~Cho, H.~Ma, W.~Xia, Y.~Yang, Z.~Liu, Z.~Huang, Z.~Jiang, X.~Lu, J.~Liu,
  Z.~Liu \emph{et~al.}, ``Emergence of new van {H}ove singularities in the
  charge density wave state of a topological kagome metal {RbV$_3$Sb$_5$},''
  \emph{Physical Review Letters}, vol. 127, no.~23, p. 236401, 2021.

\bibitem{GdMnSn_Liu}
Z.~Liu, N.~Zhao, M.~Li, Q.~Yin, Q.~Wang, Z.~Liu, D.~Shen, Y.~Huang, H.~Lei,
  K.~Liu \emph{et~al.}, ``Electronic correlation effects in the kagome magnet
  {GdMn$_6$Sn$_6$},'' \emph{Physical Review B}, vol. 104, no.~11, p. 115122,
  2021.

\bibitem{TbMnSn_Yin}
J.-X. Yin, W.~Ma, T.~A. Cochran, X.~Xu, S.~S. Zhang, H.-J. Tien, N.~Shumiya,
  G.~Cheng, K.~Jiang, B.~Lian \emph{et~al.}, ``Quantum-limit {C}hern
  topological magnetism in {TbMn$_6$Sn$_6$},'' \emph{Nature}, vol. 583, no.
  7817, pp. 533--536, 2020.

\bibitem{R166_SJia}
W.~Ma, X.~Xu, J.-X. Yin, H.~Yang, H.~Zhou, Z.-J. Cheng, Y.~Huang, Z.~Qu,
  F.~Wang, M.~Z. Hasan \emph{et~al.}, ``Rare earth engineering in
  {$R$Mn$_6$Sn$_6$} ({$R$ = Gd-Tm, Lu}) topological kagome magnets,''
  \emph{Physical Review Letters}, vol. 126, no.~24, p. 246602, 2021.

\bibitem{R166_Gu}
X.~Gu, C.~Chen, W.~Wei, L.~Gao, J.~Liu, X.~Du, D.~Pei, J.~Zhou, R.~Xu, Z.~Yin
  \emph{et~al.}, ``Robust kagome electronic structure in the topological
  quantum magnets {$X$Mn$_6$Sn$_6$} ({$X$ = Dy, Tb, Gd, Y}),'' \emph{Physical
  Review B}, vol. 105, no.~15, p. 155108, 2022.

\bibitem{Gd166_Ronning}
T.~Asaba, S.~M. Thomas, M.~Curtis, J.~D. Thompson, E.~D. Bauer, and F.~Ronning,
  ``Anomalous {H}all effect in the kagome ferrimagnet {GdMn$_6$Sn$_6$},''
  \emph{Physical Review B}, vol. 101, no.~17, p. 174415, 2020.

\bibitem{Tb166_connor}
D.~C. Jones, S.~Das, H.~Bhandari, X.~Liu, P.~Siegfried, M.~P. Ghimire, S.~S.
  Tsirkin, I.~Mazin, and N.~J. Ghimire, ``Origin of spin reorientation and
  intrinsic anomalous {H}all effect in the kagome ferrimagnet
  {TbMn$_6$Sn$_6$},'' \emph{arXiv preprint arXiv:2203.17246}, 2022.

\bibitem{V166_JFHe}
S.~Peng, Y.~Han, G.~Pokharel, J.~Shen, Z.~Li, M.~Hashimoto, D.~Lu, B.~R. Ortiz,
  Y.~Luo, H.~Li \emph{et~al.}, ``Realizing kagome band structure in
  two-dimensional kagome surface states of {$R$V$_6$Sn$_6$} ({$R$ = Gd, Ho}),''
  \emph{Physical Review Letters}, vol. 127, no.~26, p. 266401, 2021.

\bibitem{V166_MShi}
Y.~Hu, X.~Wu, Y.~Yang, S.~Gao, N.~C. Plumb, A.~P. Schnyder, W.~Xie, J.~Ma, and
  M.~Shi, ``Tunable topological {D}irac surface states and van {H}ove
  singularities in kagome metal {GdV$_6$Sn$_6$},'' \emph{Science Advances},
  vol.~8, no.~38, p. eadd2024, 2022.

\bibitem{TbVS}
E.~Rosenberg, J.~M. DeStefano, Y.~Guo, J.~S. Oh, M.~Hashimoto, D.~Lu, R.~J.
  Birgeneau, Y.~Lee, L.~Ke, M.~Yi \emph{et~al.}, ``Uniaxial ferromagnetism in
  the kagome metal {TbV$_6$Sn$_6$},'' \emph{Physical Review B}, vol. 106,
  no.~11, p. 115139, 2022.

\bibitem{V166_NP}
D.~Di~Sante, C.~Bigi, P.~Eck, S.~Enzner, A.~Consiglio, G.~Pokharel, P.~Carrara,
  P.~Orgiani, V.~Polewczyk, J.~Fujii \emph{et~al.}, ``Flat band separation and
  robust spin {B}erry curvature in bilayer kagome metals,'' \emph{Nature
  Physics}, pp. Online, https://doi.org/10.1038/s41\,567--023--02\,053--z,
  2023.

\bibitem{ScVS_1}
H.~W.~S. Arachchige, W.~R. Meier, M.~Marshall, T.~Matsuoka, R.~Xue, M.~A.
  McGuire, R.~P. Hermann, H.~Cao, and D.~Mandrus, ``Charge density wave in
  kagome lattice intermetallic {ScV$_6$Sn$_6$},'' \emph{Physical Review
  Letters}, vol. 129, no.~21, p. 216402, 2022.

\bibitem{ScVS_2}
S.~Cheng, Z.~Ren, H.~Li, J.~Oh, H.~Tan, G.~Pokharel, J.~M. DeStefano,
  E.~Rosenberg, Y.~Guo, Y.~Zhang \emph{et~al.}, ``Nanoscale visualization and
  spectral fingerprints of the charge order in {ScV$_6$Sn$_6$} distinct from
  other kagome metals,'' \emph{arXiv preprint arXiv:2302.12227}, 2023.

\bibitem{ScVS_3}
T.~Hu, H.~Pi, S.~Xu, L.~Yue, Q.~Wu, Q.~Liu, S.~Zhang, R.~Li, X.~Zhou, J.~Yuan
  \emph{et~al.}, ``Optical spectroscopy and band structure calculations of the
  structural phase transition in the vanadium-based kagome metal
  {ScV$_6$Sn$_6$},'' \emph{Physical Review B}, vol. 107, no.~16, p. 165119,
  2023.

\bibitem{ScVS_4}
X.~Zhang, J.~Hou, W.~Xia, Z.~Xu, P.~Yang, A.~Wang, Z.~Liu, J.~Shen, H.~Zhang,
  X.~Dong \emph{et~al.}, ``Destabilization of the charge density wave and the
  absence of superconductivity in {ScV$_6$Sn$_6$} under high pressures up to 11
  gpa,'' \emph{Materials}, vol.~15, no.~20, p. 7372, 2022.

\bibitem{ScVS_5}
Y.~Hu, J.~Ma, Y.~Li, D.~J. Gawryluk, T.~Hu, J.~Teyssier, V.~Multian, Z.~Yin,
  Y.~Jiang, S.~Xu \emph{et~al.}, ``Phonon promoted charge density wave in
  topological kagome metal {ScV$_6$Sn$_6$},'' \emph{arXiv preprint
  arXiv:2304.06431}, 2023.

\bibitem{ScVS_6}
S.-H. Kang, H.~Li, W.~R. Meier, J.~W. Villanova, S.~Hus, H.~Jeon, H.~W.~S.
  Arachchige, Q.~Lu, Z.~Gai, J.~Denlinger \emph{et~al.}, ``Emergence of a new
  band and the {L}ifshitz transition in kagome metal {ScV$_6$ Sn$_6$} with
  charge density wave,'' \emph{arXiv preprint arXiv:2302.14041}, 2023.

\bibitem{ScVS_7}
S.~Lee, C.~Won, J.~Kim, J.~Yoo, S.~Park, J.~Denlinger, C.~Jozwiak, A.~Bostwick,
  E.~Rotenberg, R.~Comin \emph{et~al.}, ``Nature of charge density wave in
  kagome metal {ScV$_6$Sn$_6$},'' \emph{arXiv preprint arXiv:2304.11820}, 2023.

\bibitem{ScVS_8}
S.~Mozaffari, W.~R. Meier, R.~P. Madhogaria, S.-H. Kang, J.~W. Villanova,
  H.~W.~S. Arachchige, G.~Zheng, Y.~Zhu, K.-W. Chen, K.~Jenkins \emph{et~al.},
  ``Universal sublinear resistivity in vanadium kagome materials hosting charge
  density waves,'' \emph{arXiv preprint arXiv:2305.02393}, 2023.

\bibitem{R166_romaka}
L.~Romaka, Y.~Stadnyk, V.~Romaka, P.~Demchenko, M.~Stadnyshyn, and M.~Konyk,
  ``Peculiarities of component interaction in {Gd, Er}-{V}-{Sn} {T}ernary
  systems at 870{K} and crystal structure of {$R$V$_6$Sn$_6$} stannides,''
  \emph{Journal of Alloys and Compounds}, vol. 509, no.~36, pp. 8862--8869,
  2011.

\bibitem{R166_lee}
J.~Lee and E.~Mun, ``Anisotropic magnetic property of single crystals
  {RV$_6$Sn$_6$} (r= y, gd-tm, lu),'' \emph{Physical Review Materials}, vol.~6,
  no.~8, p. 083401, 2022.

\bibitem{sun2020_03U}
Z.~Sun, Z.~Liu, Z.~Liu, W.~Liu, F.~Zhang, D.~Shen, M.~Ye, and S.~Qiao,
  ``Performance of the {BL03U} beamline at {SSRF},'' \emph{Journal of
  Synchrotron Radiation}, vol.~27, no.~5, pp. 1388--1394, 2020.

\bibitem{yang2021_03U}
Y.-C. Yang, Z.-T. Liu, J.-S. Liu, Z.-H. Liu, W.-L. Liu, X.-L. Lu, H.-P. Mei,
  A.~Li, M.~Ye, S.~Qiao \emph{et~al.}, ``High-resolution {ARPES} endstation for
  in situ electronic structure investigations at {SSRF},'' \emph{Nuclear
  Science and Techniques}, vol.~32, no.~3, pp. 1--13, 2021.

\bibitem{DFT_1}
P.~E. Bl{\"o}chl, ``Projector augmented-wave method,'' \emph{Physical Review
  B}, vol.~50, no.~24, p. 17953, 1994.

\bibitem{DFT_2}
G.~Kresse and D.~Joubert, ``From ultrasoft pseudopotentials to the projector
  augmented-wave method,'' \emph{Physical Review B}, vol.~59, no.~3, p. 1758,
  1999.

\bibitem{DFT_3}
G.~Kresse and J.~Hafner, ``Ab initio molecular dynamics for liquid metals,''
  \emph{Physical Review B}, vol.~47, no.~1, p. 558, 1993.

\bibitem{DFT_4}
G.~Kresse and J.~Furthm{\"u}ller, ``Efficiency of ab-initio total energy
  calculations for metals and semiconductors using a plane-wave basis set,''
  \emph{Computational Materials Science}, vol.~6, no.~1, pp. 15--50, 1996.

\bibitem{DFT_5}
------, ``Efficient iterative schemes for ab initio total-energy calculations
  using a plane-wave basis set,'' \emph{Physical Review B}, vol.~54, no.~16, p.
  11169, 1996.

\bibitem{DFT_6}
J.~P. Perdew, K.~Burke, and M.~Ernzerhof, ``Generalized gradient approximation
  made simple,'' \emph{Physical Review Letters}, vol.~77, no.~18, p. 3865,
  1996.

\bibitem{DFT_7}
Q.~Wu, S.~Zhang, H.-F. Song, M.~Troyer, and A.~A. Soluyanov, ``{WannierTools}:
  {An} open-source software package for novel topological materials,''
  \emph{Computer Physics Communications}, vol. 224, pp. 405--416, 2018.

\bibitem{DFT_8}
A.~A. Mostofi, J.~R. Yates, G.~Pizzi, Y.-S. Lee, I.~Souza, D.~Vanderbilt, and
  N.~Marzari, ``An updated version of wannier90: {A} tool for obtaining
  maximally-localised {W}annier functions,'' \emph{Computer Physics
  Communications}, vol. 185, no.~8, pp. 2309--2310, 2014.

\bibitem{DFT_9}
N.~Marzari, A.~A. Mostofi, J.~R. Yates, I.~Souza, and D.~Vanderbilt,
  ``Maximally localized {W}annier functions: {T}heory and applications,''
  \emph{Reviews of Modern Physics}, vol.~84, no.~4, p. 1419, 2012.

\bibitem{T166_pokharel}
G.~Pokharel, B.~Ortiz, J.~Chamorro, P.~Sarte, L.~Kautzsch, G.~Wu, J.~Ruff, and
  S.~D. Wilson, ``Highly anisotropic magnetism in the vanadium-based kagome
  metal {TbV$_6$Sn$_6$},'' \emph{Physical Review Materials}, vol.~6, no.~10, p.
  104202, 2022.

\bibitem{CoSn_kang}
M.~Kang, S.~Fang, L.~Ye, H.~C. Po, J.~Denlinger, C.~Jozwiak, A.~Bostwick,
  E.~Rotenberg, E.~Kaxiras, J.~G. Checkelsky \emph{et~al.}, ``Topological flat
  bands in frustrated kagome lattice {CoSn},'' \emph{Nature communications},
  vol.~11, no.~1, p. 4004, 2020.

\bibitem{PPS_park}
S.~Park, S.~Kang, H.~Kim, K.~H. Lee, P.~Kim, S.~Sim, N.~Lee, B.~Karuppannan,
  J.~Kim, J.~Kim \emph{et~al.}, ``Kagome van-der-waals {Pd$_3$P$_2$S$_8$} with
  flat band,'' \emph{Scientific Reports}, vol.~10, no.~1, p. 20998, 2020.

\bibitem{Y166_yang}
T.~Yang, Q.~Wan, Y.~Wang, M.~Song, J.~Tang, Z.~Wang, H.~Lv, N.~Plumb,
  M.~Radovic, G.~Wang \emph{et~al.}, ``Evidence of orbit-selective electronic
  kagome lattice with plane flat band in associated paramagnetic
  {YCr$_6$Ge$_6$},'' \emph{arXiv preprint arXiv:1906.07140}, 2019.

\bibitem{liu_AlFeB}
Z.~Liu, S.~Yang, H.~Su, L.~Zhou, X.~Lu, Z.~Liu, J.~Gao, Y.~Huang, D.~Shen,
  Y.~Guo \emph{et~al.}, ``Non-fermi-liquid behavior and saddlelike flat band in
  the layered ferromagnet {AlFe$_2$B$_2$},'' \emph{Physical Review B}, vol.
  101, no.~24, p. 245129, 2020.

\bibitem{NFCA_liu}
Z.-H. Liu, P.~Richard, K.~Nakayama, G.-F. Chen, S.~Dong, J.-B. He, D.-M. Wang,
  T.-L. Xia, K.~Umezawa, T.~Kawahara \emph{et~al.}, ``Unconventional
  superconducting gap in {NaFe$_{0.95}$Co$_{0.05}$As} observed by
  angle-resolved photoemission spectroscopy,'' \emph{Physical Review B},
  vol.~84, no.~6, p. 064519, 2011.

\end{thebibliography}

\end{document}